\documentclass[cits]{PoS}

\title{Asteroseismology of stars on the upper main sequence}
\ShortTitle{Asteroseismology of upper main sequence stars}

\author{\speaker{Patrick Lenz}
  \thanks{The author is partially supported by the Polish MNiSW grant No.
N N203 379636.}
  \\
  Copernicus Astronomical Center, Polish Academy of Sciences, ul. Bartycka 18, 00-716 Warsaw, Poland\\
  E-mail: \email{lenz@camk.edu.pl}}

\abstract{I review the properties of pulsators located on the upper main sequence in the HR diagram and discuss asteroseismic inferences on the internal structure of stars of spectral type A and B. Special attention is given to the problem of uncertainties in stellar opacities in modelling.}

\FullConference{Frank N. Bash Symposium New Horizons In Astronomy,\\
		October 9-11, 2011\\
		Austin Texas}

\begin{document}

Pulsating stars on the upper main sequence in the HR diagram commonly exhibit a convective core (which appears if M > 1.2 M$_\odot$) and a radiative envelope with thin convection zones close to the surface. The evolutionary status of pulsators located in this region in the HR diagram can be manifold, however: beside common main sequence pulsators (in which hydrogen core burning takes place) we also have pre-main sequence stars (no efficient nuclear reactions) and post-main sequence stars (hydrogen shell burning). We distinguish different types of pulsators along the main sequence band:
among B and late type O stars we have the so-called $\beta$~Cephei pulsators with periods of hours and masses of 8--20~M$_\odot$ 
%(mention Be stars?)
and the long-period SPB oscillators (slowly pulsating B stars) with periods of days and masses of 3--12~M$_\odot$.
Moving towards lower masses, there are the $\delta$~Scuti pulsators (M = 1.5--2.5~M$_\odot$), which are dwarfs or giants of spectral type A2--F5 located in the extension of the Cepheid instability strip with periods of 0.02--0.3d. Pulsating magnetic stars among A stars are known as roAp pulsators with periods of 5--15 minutes. Among F-type stars there are the $\gamma$~Dor pulsators with masses of 1.4--1.6~M$_\odot$ and periods of 0.3--3~d.
Recent observational reviews on these pulsators based on satellite photometry can be found in \cite{2011uytterhoevenI,2011balonaIII,2011balonaguzik} for A and F pulsators and for B stars in \cite{2011balpig,2009degroote}.
Prior to the discussion of recent asteroseismic results on the physics in these stars I will review some basics of stellar pulsation.

%___________________________________________________________________________
\section{Stellar oscillations}

Stellar pulsation occurs if a star undergoes free or forced oscillations. In the limit of slow rotation, the star is spherically symmetric in its equilibrium. In this case the geometrical perturbation of the equilibrium through a pulsation mode can be characterized by a spherical harmonic $Y_\ell^m(\theta,\phi)$, where $\theta$ is the colatitude, $\phi$ the azimuthal angle, $\ell$ the degree (i.e. the number of nodal lines on the surface) and $m$ the number of nodal lines crossing the equator. If $\ell=0$ we have radial pulsation (i.e.~a spherically symmetric oscillation) and if $\ell>0$ we have nonradial oscillations. For a spherical degree, $\ell$, the eigenfunctions are degenerate by (2$\ell$+1)-folds in $m$. Rotation lifts this degeneracy and leads to mode frequencies depending on $m$. 
Finally, a pulsation mode is also characterized by the number of nodes, $n$, in the radial component of the displacement between the center and the surface. The radial fundamental mode has no node, i.e. $n$=0, the first overtone $n$=1, and so forth.

The symmetry axis of pulsation is commonly aligned to the dominant symmetry axis in the star. It is often the rotation axis; however, in case of the presence of a strong magnetic field it is the magnetic axis, or in a close binary system the tidal axis. If no symmetry axis clearly dominates, the pulsation symmetry axis may lie in between two symmetry axes, e.g., in between the rotation axis and the magnetic field axis \cite{2002bigot}. Such an example has already been found in a magnetic A-type star observed by the {\it Kepler} satellite \cite{2011kurtz}.

\subsection{Propagation of waves inside a star}

As for every oscillating body its structure and composition determines its frequencies. The solution of the oscillation equations for a star reveals two characteristic (critical) frequencies:
(i) the Lamb frequency, $L_\ell = \sqrt{\ell (\ell+1)} \frac{c_s}{r}$, where $c_s$ is the local speed of sound, and 
%unno: frequency scale of one horizontal wave length divided by the local sound speed, unno.11: sound wave travels a wavelength 2pir/l horizontally in the period 2pi/$L_\ell$
(ii) the Brunt-V\"ais\"al\"a frequency, $N$, defined as $N^2 = g \left( \frac{1}{\Gamma_1} \frac{d \ln p_0}{dr} - \frac{d \ln \rho_0}{dr} \right)$, where the subscript 0 denotes equilibrium values.
 The Lamb frequency corresponds to the inverse travel time of a sound wave, i.e. a wave front propagates the distance $2\pi r/\ell$ horizontally within the period $2\pi/L_\ell$. The Brunt-V\"ais\"al\"a frequency describes the frequency of the adiabatic oscillation of a bubble of gas in vertical direction under the influence of buoyancy. %(gravity)
 Both critical frequencies depend on the local physical conditions inside the star and they determine the cavities in which oscillation may take place. The propagation zone for acoustic waves (p modes) is defined by $\sigma > L_\ell$ and $\sigma > N$, where $\sigma$ is the oscillation frequency, and the gravity wave (g mode) propagation zone resides in regions where $\sigma < L_\ell$ and $\sigma < N$. In between these two cavities the amplitude of an oscillation mode decreases exponentially with distance, this region is therefore called evanescent zone. Additionally, at high frequencies the acoustic cavity is limited by the acoustic cut-off frequency, above which the oscillation is no longer reflected at the outer boundary but propagates outwards in the atmosphere.

%This basically means that acoustic modes (restoring force is pressure) have high frequencies while gravity modes (restoring force buoyancy) exhibit low frequencies.

\begin{figure}
  \centering
  \includegraphics[width=\textwidth]{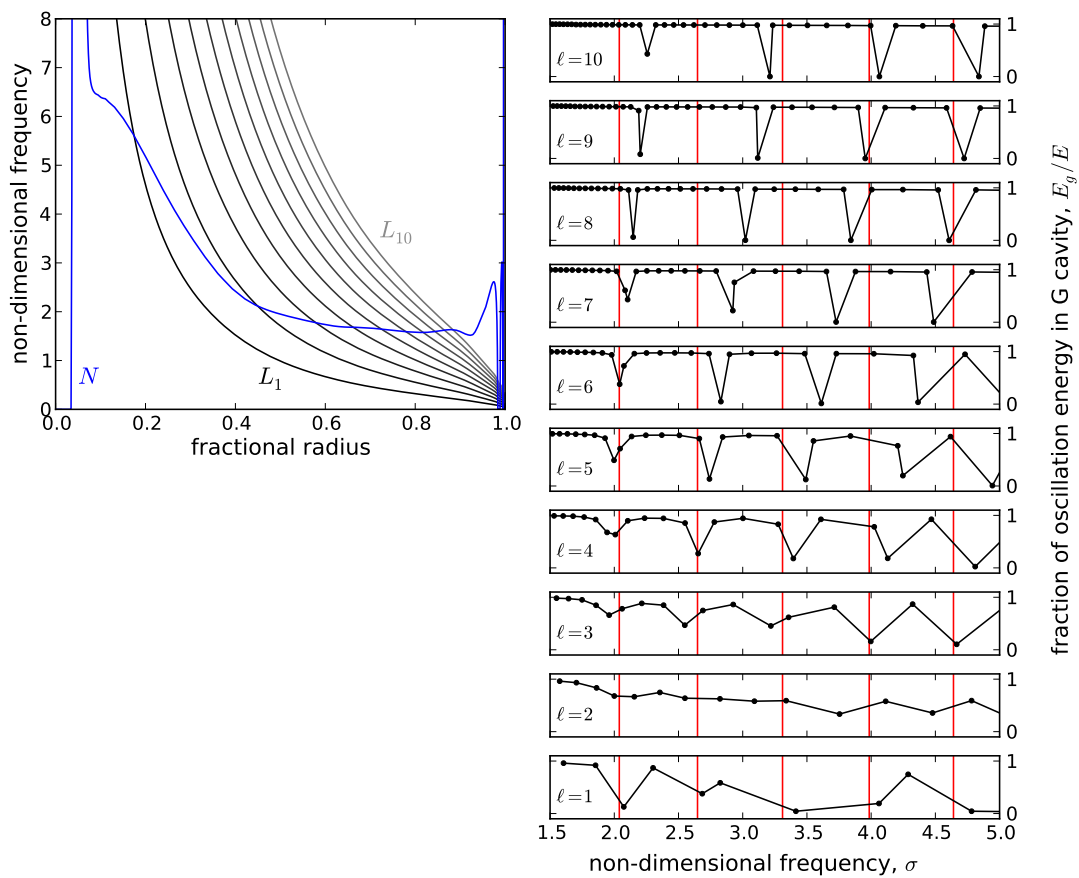}
  \caption{Left panel: propagation diagram for modes up to $\ell$=10 in a model of the $\delta$~Scuti star 44~Tau. Right panel: fraction of oscillation mode energy confined in the g~mode propagation zone for modes of different $\ell$. Modes with $E_g/E \approx 1$ are effectively trapped in the g-mode cavity, while modes with $E_g/E\approx 0$ are trapped in the envelope. The location of the radial modes is indicated by red vertical lines.}
  \label{fig:prophl}
\end{figure}

A typical example of propagation zones in a pulsating star of $\approx$~1.9~M$_\odot$ at the end of hydrogen core burning is given in the left panel of Fig.~\ref{fig:prophl}. The high values of $N$ close to the center are due to the strong gradient in mean molecular weight which grows during main sequence evolution because of the receding convective core. This development allows g~modes to move to higher frequencies. In the outer envelope two convection zones ($N<0$) corresponding to partial ionization of He II and HeI/H are located. A significant part of the energy in these zones is still transported by radiation, but the efficiency of convection increases with decreasing effective temperatures of the star. If the two wave propagation zones are separated only by a small evanescent zone a ``tunnel effect'' may occur. If the evanescent zone is large enough, however, the propagation zones can be treated as independent and oscillation energy is effectively trapped in a given propagation zone. If this is the case we speak of \emph{trapped modes}. 
%In fact the evanescent zone can be seen as a potential wall (lt. unno). 
% boundary condition may be applied in the evanescent zone
It can be seen in Fig.~\ref{fig:prophl} that while we expect oscillation modes of higher $\ell$ to be effectively trapped in the envelope or in the interior; for low degree modes only partial trapping occurs. In fact partial trapping is least effective for $\ell=2$ modes in main sequence stars because for these modes the evanescent zone is thinnest in the range of typically excited modes and therefore these modes are strongly coupled to the interior. Consequently, they have both g~mode and p~mode properties and are therefore called \emph{mixed modes}. This effect is also illustrated in the right panel of Fig.~\ref{fig:prophl} which shows the fraction of oscillation energy of a mode confined in the gravity wave propagation zone, $E_g/E$.
Modes trapped in the envelope, i.e. modes with low $E_g/E$, are essentially decoupled from the interior and have the highest probability to be observed.
The agglomeration of modes trapped in the envelope close to the frequency of the fundamental radial mode is due to the fact that the modes of all $\ell$ values are limited by the Brunt-V\"ais\"al\"a frequency which itself is $\ell$-independent (see left panel in Fig.~\ref{fig:prophl}).
Although modes with high spherical degrees suffer from stronger cancellation effects compared to $\ell$=1,2 modes \cite{1999balona} they are now also detectable with present day high-precision satellite photometry.

%evolved stars: 'non-radial mode trapping becomes effective at lower ell values than in main-sequence stars (connected with strong damping in the G zone, found: WAD1977a, Osaki1977, recently discussed: vanHoolst1998)'
%'large values of N in interior of He burning stars imply very short radial wavelengths of the eifenfunctions, hence large radiative losses' trapped modes = unstable modes and have no adiabatic counterpart

%wad: Why with rising frequency, dipolar modes move from vicinity of radial mode toward the the mid position between consecutive radial mode is easy to understand. At low frequencie the acoustic cavity for dipolar modes is limited, like for radial modes,  by the the B-V frequency, which is l-independent while at high frequencies it is the Lamb frequency.\\

%\newpage
\section{Excitation and damping of oscillation modes}

What causes an oscillation mode to grow in amplitude? We distinguish between free and forced oscillations. In the latter case a linearly damped oscillation is excited by a periodic external force, e.g., due to a periodic tidal distortion, or generally, due to resonance.
In case of free oscillations the oscillations are excited by an internal driving mechanism. Such an excitation mechanism should be located in a region that lies within a propagation zone and the propagation zones should exclude the common damping regions in a star. Moreover, the oscillation mode should not exhibit a node in the driving region. The most relevant mode driving mechanism for self-excited oscillation are outlined in the following paragraphs.

\subsection{$\kappa$-mechanism}

\begin{figure}[h!]
  \centering
  \includegraphics[width=.5\textwidth,bb= 0 30 630 800]{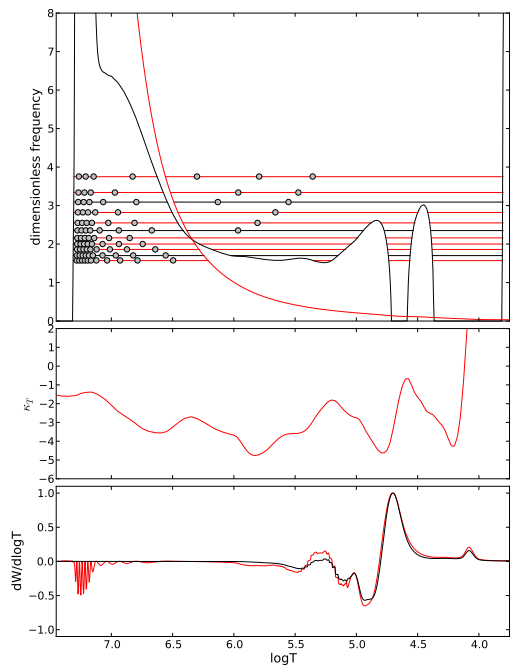}
  \caption{Propagation diagram, temperature derivative of opacity, $\kappa_T$, and differential work integral for two $\ell=2$ modes (p~mode in black, g~mode in red) as a function of $\log T$ in a 44~Tau model.}
  \label{fig:driving}
\end{figure}

Opacity ($\kappa$), i.e. the quantity which describes the transport of radiation through matter, is temperature-dependent and can act as a valve under certain circumstances. Depending on the layer inside the star, $\kappa$ increases or decreases with increasing temperature which is reflected in the behaviour of the temperature derivative of opacity, $\kappa_T$ (see Fig.~\ref{fig:driving}). One condition for the $\kappa$-mechanism to work in a certain region in a star is that this opacity derivative increases in the outward direction. Consequently during a compression phase radiative flux is blocked and performs work as can be seen from the differential work diagram in Fig.~\ref{fig:driving}.
% rogers-science: 'kappa-mechanism occurs in regions where the opacity rises sharply with increasing temperature in ionization zones. Under such conditions, a contraction of the star can increase the opacity in this region, trapping the energy flux emanating from the interior.'

The conditions for pulsational instability of a mode (i.e. for its amplitude to grow with time) was reviewed, e.g., by \cite{1999pamyatnykh} and only the main points will be repeated here: 
(i) the amplitude of the pressure eigenfunction has to be large and vary slowly within the driving zone,
(ii) the pulsation has to occur faster than the redistribution of thermal energy, i.e. the thermal timescale in the driving region has to be comparable or longer than the period of the mode. 
 The second condition determines the depth of the driving zone in a star.
The given conditions can be met by two frequency ranges in the same star, as for example in hybrid B pulsators: low-order low-degree acoustic or mixed modes with periods of  3--6 hours ($\beta$~Cep-type pulsation) and high-order low-degree gravity modes with periods of  1.5--3 days (SPB-type pulsation).
Condition (ii) implies that these longer period modes are driven in slightly deeper layers than the $\beta$~Cep-type modes.
%A similar ``instability bump'' for high-order gravity modes of higher degree occurs in F stars in the classical instability strip, however, convection is more effective in these objects and needs to be considered.

%\textit{recent studies: g-mode excitation: massive post-MS stars: shell conv. zone + radiative core\\ pre-MS: no core damping, therefore: pre-MS SPB? low-degree, high-order in intermediate stars are excited}

There are also important damping effects such as radiative dissipation in the gravity mode cavity. The strength of this effect depends on the efficiency of mode trapping in this cavity and it is therefore strong for g~modes (see lower panel of Fig.~\ref{fig:driving}).
The dissipation increases with the degree of central condensation, which generally increases with age. E.g., Cepheids have a very high central concentration and the strong radiative dissipation in the core region damps g~modes before they are reflected at the center. Hence in these stars only acoustic modes trapped in the envelope are observed. In $\delta$~Scuti stars radiative dissipation of modes provides strong damping for g~modes of higher $\ell$. At $\ell>12$ only acoustic modes trapped in the envelope remain unstable \cite{1999balona}.

%maybe this also occurs in HADS (less centrally condensed as CEPH, but more centrally condensed than ZAMS DSCT), more evolved but this still has to be shown.

%less centrally condensed structure, radiative dissipation in core not enough to inhibit existence of standing modes of gravity waves, whole star as oscillating unit, especially for low l modes. DSCT more centrally condensed than zams stars, gravity waves have short wavelengths in the core. eigenfunc of nonradial mode have a lot of nodes.\\

%U.Lee: dissipation occurring in the g-zone overcomes excitation of the kappa-mechanism in the envelope for the nonradial mode in case of l=3, but does not stabilize the mode with l=1\\

%inertia of p-modes much smaller than g-modes

\subsection{Convective driving and convective blocking}

Efficient convection may also cause self-excited oscillations. For example in ZZ~Ceti stars the convective energy flux in the H-ionization zone is much larger than radiative flux. Since convection occurs on a much shorter timescale compared to g~mode oscillations, the instantaneous adjustment of convection to pulsation leads to thermal energy being stored in the convection zone during the compression phase of pulsation (i.e. the convection zone stores heat) thus providing efficient driving \cite{1999goldreich}.

However, excitation of g~mode pulsation is also possible if the convective time scale is very long against the pulsation period, such as in F stars in case of $\gamma$~Dor pulsation. The radiative flux may be effectively blocked by convection at the base of the convection zone, which also leads to heating upon compression \cite{2000guzik}. For this mechanism to work, the base of the convection zone should match the region where the thermal relaxation time is similar to the pulsation periods of the modes \cite{2011balonaguzik}.

%Recent observations of GDOR pulsators reveal mostly only one or two modes \cite{2011balonaI}.

%convection:\\
%Kupka \& Montgomery: small alpha in H-zone, high alpha in He-zone\\
%Trampedach

%GDOR: from Balona, Guzik et al. (2011): It is the location of the base of the convective envelope which seems to be the key element in driving γ Dor pulsations. In the mid-dle of the instability strip, the base is located in the transition region where the thermal relaxation time roughly matches the pulsation periods. Driving is therefore very efficient. For hotter stars, the con-vective envelope is thinner which means that the thermal relaxation time is shorter than the pulsation period. The heat capacity is small and driving is inefficient. Since the size of the convection zone is determined by the adopted mixing length, the resulting blue edge is sensitive to the adopted mixing length, α (Dupret et al. 2004). The stability of models at the red edge has a different origin. For the cooler models, radiative damping of g modes overcomes driving by convective blocking and stabilizes the model. Good agreement with ground-based observations is found for α = 2. For α = 1.5 and lower, the calculated instability strips do not match the observations at all (Dupret et al. 2004). some kappa mech. driving from Fe bump

\subsection{Stochastic driving}

Forced excitation of modes within a certain frequency band due to acoustic noise in an efficient convective envelope is another possible driving mechanism.
This excitation mechanism drives the 5-min oscillations in the Sun. Although envelope convection in upper main sequence stars is less efficient than in the Sun, evidence for stochastically excited oscillations has been found in a few stars, partly additional to common opacity mechanism driving. There are examples among A/F stars \cite{2011antoci} and O/B stars \cite{2009belkacem,2010degroote}.
%Cantiello et al. (2009): sub-surface convection models, Belkacem et al. (2010),
%, CoroT O-type star HD 46149' (Samadi?)
However, in one of these cases the stochastic nature of the modes could not be confirmed, since non-linear resonant mode excitation by the large-amplitude radial mode provides similar observational features \cite{2011aerts}.
%However, as pointed out by \cite{2011aerts} non-linear resonant mode excitation by the large-amplitude radial mode may cause similar mode features in one of these cases.
%However, in one of these cases stochastic modes could not be confirmed, since they can be explained by non-linear resonant mode excitation by the large-amplitude radial mode.

%Balona et al. 2011 MNRAS also questioned this because of lack of such oscillations in Kepler

%Corot and solar-like oscillations: Mathur et al. 2010 and references therein
%Kepler: Chaplin et al. 2011

\subsection{Modification of mode excitation through a magnetic field}

The presence of a magnetic field modifies driving by the $\kappa$-mechanism in two ways: (i) convection is inhibited in the polar regions of the magnetic field, where inward propagating magnetic slow waves carry away pulsational energy, (ii) more effective gravitational settling leading to hydrogen-enriched surface layers.
In fact, a nonadiabatic analysis \cite{2005saio} for A-type stars showed that a dipole magnetic field stabilizes low-order acoustic modes at 1 kG, while high-order modes of $\ell=1,2$ (roAp-type pulsation) become pulsationally unstable due to driving in the H-ionization zone.

%generally we have two effects: gravitational settling and/or magnetic damping:
%ersteres reduces He abundance, H-enriched surface layers

%convection suppressed in mag. polar regions (vertical mag. field), pulsation generates mag. slow waves, carry away pulsation energy (slow waves: short wavelengths, propagate inward)

%Houdek 2003: magnetic lines of force are bent by convective motion, resulting in traveling waves along the lines of force

%rodriguez-lopez 2011:
%magnetic inhibition of convection leads to models having larger radii and lower temperatures

%Sousa \& Cunha (2011) 'handler2011: first theoretically-based explanation for the diversity found observationally in the atmospheric behaviour of the oscillations of roAp stars'

%\subsection{Saturation of driving}

%not ready, maybe replace section with one sentence

%mode amplitude is limited when:
%driving mechanism is saturated (i.e., no further mechanical energy is available)
%resonant mode coupling (mechanical energy in the mode is used to drive another mode with nearly the same frequency)
%role of damping in HADS is particularly relevant to this problem.
% \cite{Balona1999}: instability may be saturated by high degree modes, multimode saturation of the instability may take the form of an unsteady (periodic or chaotic) limit cycle, in which the energy associated with radial pulsation has been transferred to non-radial modes of high degree. multimode saturation requires non-linear modelling

%__________________________________________________________________________________________________
\section{Asteroseismic inferences from upper main sequence stars}

\subsection{Probing stellar opacities}

Opacities are a fundamental ingredient in the calculation of stellar structure and evolution and, therefore, asteroseismic models are very sensitive to them. 
The opacity coefficients $\kappa$($T$,$\rho$,$X_i$) define the interacting cross sections of radiation with matter and determine the efficiency of radiative energy transport in a star. Naturally, these opacity coefficients are high for elements with many electrons. In most of the interior of a main sequence star hydrogen and helium are completely ionized except for the outer envelope. Consequently in the deeper interior heavy elements have a high contribution to the opacity coefficients. Therefore, despite their small mass fractions, heavy elements play an important role in stellar physics.

For practical reasons we commonly use tabulated Rosseland-mean opacities, $\kappa_R$, in stellar models. Two sets of opacity tables are currently widely used, the Lawrence Livermore National Laboratory opacity table computed with the OPAL code \cite{1996iglesias} and the tables of the international Opacity Project collaboration (commonly referred to as OP opacities) \cite{2005badnell}. 

In their most recent incarnations the OPAL calculations are based on the 21 most abundant elements in a star
% (H, He, C, N, O, Ne, Na, Mg, Al, Si, P, S, Cl, Ar, K, Ca, Ti, Cr, Mn, Fe, and Ni) 
while OP considers 17 species (i.e. the same elements as OPAL with exception of P, Cl, K and Ti, which have the lowest abundances in the given mixture). There are also differences in the computational approach, e.g., in the equation of state (EOS) which determines ionization equilibria and level populations required for opacity calculation. The EOS used by OP calculations \cite{1988hummer} is based on the 'chemical picture',
%relating the effect of the surrounding plasma on the internal structure of ions and atoms
 while the OPAL EOS \cite{2002rogers} considers a 'physical picture'.
%, taking into account only an ensemble of electrons and nuclei. 
A detailed comparison of both EOS is given in \cite{2006trampedach}.

\begin{figure}[h!]
  \centering
  \includegraphics{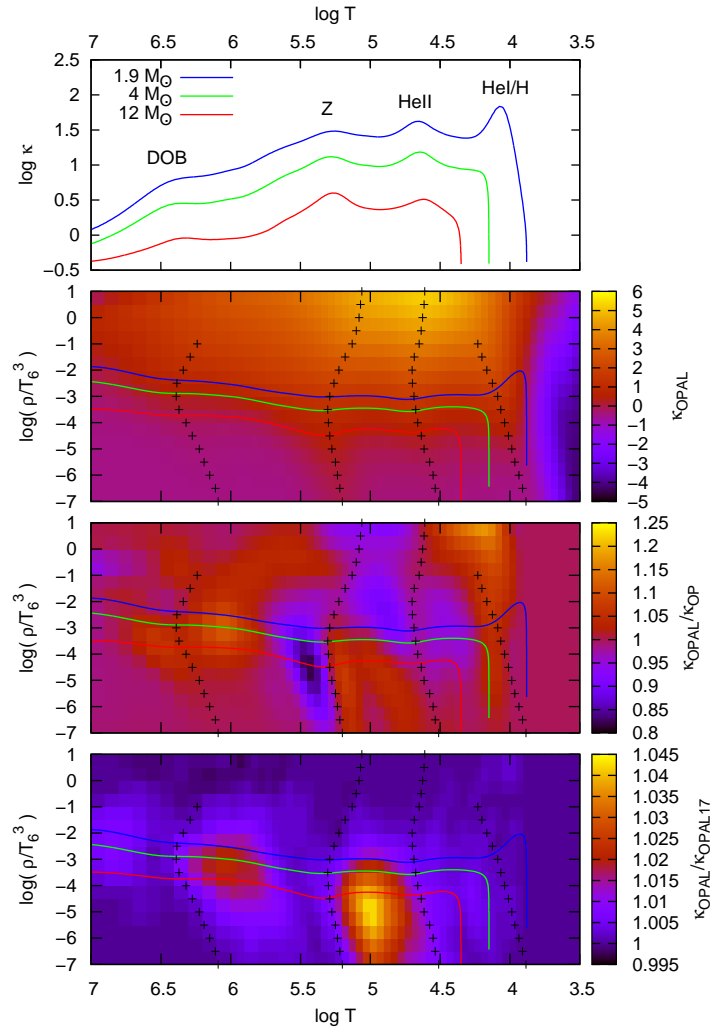}
  \caption{Probing differences in opacity tables by means of main sequence pulsators. The uppermost panel shows Rosseland mean opacity in the envelope of a $\delta$~Scuti model (1.9 M$_\odot$), a SPB-type model (4 M$_\odot$) and a $\beta$~Cephei model (12 M$_\odot$). The adjacent panel shows the profile of these models in the temperature-density plane of the OPAL opacity table for X=0.74, Z=0.0134. Around logT=3.95 the transition to the low-temperature opacities by \cite{2005ferguson} occurs. For clarity crosses roughly mark the location of the different opacity bumps. The third panel shows $\kappa_{\rm OPAL}/\kappa_{\rm OP}$ in the same plane and the lowermost panel shows the effect of the four elements (P,Cl,K,Ti) which are included in OPAL but not in OP.}
  \label{fig:opac}
\end{figure}

Seaton et al. \cite{2004seaton} compared the Rosseland mean opacities for six elements (H, He, C, O, S, Fe) and generally found good agreement between OPAL and OP. If one compares the OPAL and OP opacities for the full set of species, the differences are more significant, however. The third panel from the top in Fig.~\ref{fig:opac} shows $\kappa_{\rm OPAL}/\kappa_{\rm OP}$ evaluated for the most recent solar element mixture \cite{2009asplund} as a function of temperature and density.  The most striking difference is the well known fact that the metal opacity bump is shifted to higher temperatures in the OP data. Another feature is that the OPAL table exhibits higher opacities compared to OP at log T = 6 and log($\rho/T_6^3$) = -3  where $T_6 \equiv T/10^6$. The figure illustrates the stellar profile of different pulsators such as a $\delta$~Scuti star (1.9 M$_\odot$), a SPB pulsator (4 M$_\odot$) and a $\beta$~Cephei star (12 M$_\odot$) in the temperature-density plane of the opacity table.
As can be seen, different regions in the opacity tables are probed by pulsators on the upper main sequence and the differences between OP and OPAL influence the pulsation models for these stars. 

As discussed in the previous section a positive temperature derivative of opacity, $\kappa_T$, in outward direction is one important condition for driving pulsations. This condition is fulfilled at the hotter wings of bumps in opacity. We will now discuss the different opacity bumps as shown in the top panel of Fig.~\ref{fig:opac} and relate them to their corresponding types of pulsators:

The {\bf He~II bump} at log T $\approx$ 4.65 is due to the second ionization of helium. It is responsible for pulsation in the classical instability strip in the HR diagram, e.g., $\delta$~Scuti pulsators, RR Lyrae stars and Cepheids.

The {\bf Z bump} is formed due to a large number of intra-M shell transitions in highly excited ions of iron-group elements which take place at approximately log T $\approx$ 5.3. Iron has the strongest contribution, but also Nickel is a significant contributor of opacity, despite its lower abundance \cite{2006jefferyB}. The temperature of the Nickel opacity bump is significantly higher in the OP data compared to OPAL which poses an interesting problem.
The Z bump is responsible for pulsational driving in massive main sequence B stars ($\beta$~Cep and SPB type), but also in evolved stars such as hot subdwarf B and O stars (sdB, sdO) on the extreme horizontal branch.
%pulsating in low-radial order p and high order g mode pulsation in \cite{2008jeffery} and references therein, \cite{2003fontaine}, \cite{2006jefferyA} or other extreme helium stars \cite{1993saio}.
 The Z bump instability strip for radial modes connecting main sequence pulsators of $\beta$~Cep and SPB type with low mass sdO and sdB pulsators is shown, e.g., in Fig.~1 in \cite{2008jeffery}.

The {\bf deep opacity bump (DOB)} occurs from the partial ionization of L-shell electrons of Iron at log T $\approx$ 6.3 and of K-shell electrons of C, O, Ne at log T $\approx$ 6.2. %This bump is important for helioseismology.
%, and 'lithium depletion in low-mass stars'. 
This bump may excite pulsation 
%of low-degree g modes of intermediate radial order 
in hydrogen-rich Wolf-Rayet stars \cite{2006townsend} and GW Vir stars \cite{2005gautschy}.
%However, we could also have strange mode pulsation mechanism here: 2008Glatzel, 2009Glatzel

Asteroseismology probes the stellar opacities through different observables for each oscillation mode: \\
(i) its pulsational instability (in opacity-driven pulsators): as instability critically depends on the opacity profile in the vicinity of the bump responsible for driving (see Fig.~\ref{fig:driving}) \cite{2008montalban}\\
(ii) its frequency: because the oscillation frequencies depend on the radiative properties of the stellar medium in the propagation zone(s) of the mode.\\
%, damping in convection zones. opac determines density distribution. This will affect modes with antinodes where the differences are. Changing opacities shift model when fitting frequencies to observed values: shift in HRD indirectly modifies opacities\\
(iii) its nonadiabatic $f$ parameter (i.e. the ratio of bolometric flux perturbation to the radial displacement at photosphere level): because the complex quantity $f$ probes the nonadiabatic regions and, therefore, is sensitive to the conditions in regions with temperatures below the temperature of the driving opacity bump. Seismic analyses involving the $f$~parameter are commonly termed complex asteroseismology \cite{2010jdd,2009jdd}.

History has taught us that updates in theoretical opacity calculations (due to improved physics and consideration of additional elements) generally lead to an increase in opacity and an improvement of our understanding of the excitation of pulsation in stars. For example, based on observational evidence for higher opacity, Simon pleaded for a reinvestigation of heavy element opacities in 1982 \cite{1982simon}. In 1992 the OPAL team published new tables which included new spin-orbit interactions causing an increase of opacity at the Z bump. This led to the successful explanation of the driving of the observed pulsation in B stars \cite{1993wadpam,1993wadmospam}.
%{1992cox,1992moskalik,1992kiriakidis,1993wadpam,1993wadmospam,1993gautschy}
 Consequently, asteroseismology can serve as a tool to help us to identify flaws in certain parts of the opacity tables. Today there is again observational evidence that present day mean opacities underestimate real opacity.

%In the following discussion of case studies these effects are not disentangled.

Opacities have been tested by conducting instability surveys (i.e. calculation of instability strips and comparing them to observed positions of pulsators in the HRD) for different types of pulsators. Various effects on the location of the instability strips of upper main sequence pulsators are discussed in \cite{1999pamyatnykh} and the effect of the latest update of OP opacities and solar element abundances in \cite{2007miglio,2008zdravkov}. Generally, the last update of OP data in 2005 resulted in larger instability domains for both $\beta$~Cephei and SPB pulsators in the HR diagram, and the domains shifted to hotter temperatures. 
%The differences between OPAL and OP at the Z bump are getting less pronounced with lower mass which is clearly seen in Fig.~\ref{fig:opac}..
%The use of new solar element abundances as shown in AAP2007 is that 'at fixed Z new Fe group abundances higher than the older ones, because Z value mainly determined by CNO,Ne'. 
 The overlap region that hosts $\beta$~Cephei and SPB hybrid stars is very sensitive to these changes and consequently a good observational probe. For this reason many of these hybrid pulsators were observed in detail during the recent years. Particularly well-studied cases are
12 Lac (M $\approx$ 11.5 M$_\odot$, \cite{2006handler,2008wadaap,2009desmet}), 
$\nu$ Eri (M $\approx$ 9.5 M$_\odot$, \cite{2004handler,2004aerts,2004deridder,2004pamyatnykh,2005jerzy,2008wadaap}) 
and
$\gamma$ Peg (M $\approx$ 8.5 M$_\odot$, \cite{2006chapellier,2009handler,2009handlerII,2009zdravkov,2011pandey}).
It is striking that in asteroseismic models of these stars certain common problems occur. With OPAL the predicted frequency range of unstable $\beta$~Cephei modes is too narrow. The use of OP data improves the predicted mode instability for $\beta$~Cephei modes in comparison with OPAL but additional instability is needed. With OPAL it is often not possible to excite low-frequency SPB modes at all, while with OP we do obtain unstable high-order g-modes but matching them with observed frequencies is difficult \cite{2008wadaap}.
For example in $\gamma$ Peg $\ell=2$ SPB modes are predicted to be excited with OP opacities but observations indicate $\ell=1$ \cite{2009zdravkov}.

The underestimation of driving for $\beta$~Cep modes can be rectified by increasing the opacity bump responsible for driving, since driving is more effective if the bump is more prominent in comparison with its surrounding. An increase in OP opacity by about 50\% at the Z bump at log~T$\approx$5.3-5.5 and an increase of a few \% up to 20\% at the deep opacity bump at log~T$\approx$6.3 improves the agreement in terms of excitation and frequency fits for $\beta$~Cephei pulsation \cite{2008zdravkovII,2009zdravkov}. 
Additional evidence comes from pulsating B stars found in the low-metal environment SMC for which opacity enhancement is needed at the Z bump to excite the observed modes \cite{2010salmon}.

Unfortunately there are not many successfully modelled A-F stars. However, there is one star which is well understood and provided some hints on opacities: the $\delta$~Scuti pulsator 44~Tau (M $\approx$ 1.9 M$_\odot$, \cite{2007antoci,2010lenz}). For this star the fact that OP opacities are lower than OPAL by 10\% at log~T=6.05 caused serious problems in modelling, in particular when fitting the period ratio \cite{2007lenz,2008montalban}. This temperature is close to the deep opacity bump and the problem of OP opacities can be solved by an increase of opacity at log T=6.05 \cite{2010lenz} which is close to the temperature region where an opacity increase is also required in B stars.

Studies focussing on the nonadiabatic $f$~parameter, which is sensitive to opacities in the outer envelope, reveal an ambivalent picture.
For the $\beta$~Cep star $\theta$ Ophiuchi (M $\approx$ 8.2 M$_\odot$, \cite{2005handler,2005briquet,2007briquet,2010lovekin}) as well as for $\nu$ Eri \cite{2010jddwal} the comparison between emprically determined $f$ and theoretically computed values show preference for OPAL opacities \cite{2009jddwal,2010jddwal}. In $\gamma$~Peg, however, models based on both OP and OPAL fail to reproduce the nonadiabatic properties of observed SPB-type modes \cite{2010waljdd}.

%Consequently modelling the non-adiabatic can also be improved by reducing uncertainties in opacities.
%It should be noted that all these stars are slow rotators. Although this has the advantage that we can disentangle the effects of rotation to some degree, effect of diffusion may also be important.

Consequently, \cite{2008wadaap} pointed out the requirement of enhancement of opacity in the driving region in order to explain the instability of SPB-type frequencies in these hybrid pulsators.
In B stars  opacity enhancement tests have the following effect: increasing opacity at the driving bump (e.g., the Z bump in B stars) affects mode instability for SPB and $\beta$~Cep pulsation but does not change the frequencies significantly because the opacity was changed at low densities. Since the DOB is located in denser layers it influences the mode frequencies strongly and hence, if we fit theoretical frequencies to observed counterparts, the position in the HR diagram changes.
Additional evidence for more opacity also comes from recent observations hardening the claim of $\beta$~Cep-type pulsations in O stars, e.g., the O9V star HD 46202 \cite{2011briquet}, or the O9.5V star $\zeta$~Oph \cite{2005walker}. %1997kambe, 
%(compare position in hrd with instability strip in 2008zdravkov).
These observations could be explained by widening the $\beta$~Cep instability strip by means of additional opacity at the Z bump.
%\emph{Generally, the temperature of an opacity peak that drives pulsation is related to the effective temperature of the corresponding instability domain in the HRD. If that opacity bump is shifted to higher temperature the instability strip is also moved to higher effective temperatures in the HRD.}
In the Sun an opacity enhancement of 30 \% at the base of the convective zone (which approximately corresponds to the deep opacity bump) and a few percent in the solar core has also been suggested to solve the discrepancy between the solar model and helioseismology based on the current version of solar abundances \cite{2009jcd}.

A possible explanation for underestimated opacity, apart from uncertainties in the calculations (especially concerning the peak temperature of the Nickel bump), is the possible opacity contribution from elements with low abundances which currently are not included in opacity calculations. In the OPAL calculations four species (P, Cl, K, Ti) are considered which are not included in the OP computations. Using the OPAL web interface\footnote{http://opalopacity.llnl.gov/new.html} we retrieved a table which adopts the element abundances from the most recent solar mixture \cite{2009asplund} but setting the number fraction of these four elements to zero. Since this table is then based only on 17 elements we denote this table as OPAL17 hereafter. Due to the renormalization of the number fractions, we have a minor abundance increase in all metal elements. The lowest panel in Fig.~\ref{fig:opac} shows the ratio $\kappa_{\rm OPAL}/\kappa_{\rm OPAL17}$ for the solar chemical composition illustrating where these four elements contribute opacity. They augment up to a few \% of opacity for upper main-sequence stars at the hot wing of the Z bump around log~T$\approx$5 and close to the DOB at log~T$\approx$6.0. Consequently one may conclude that the inclusion of more species may partly solve the problems in asteroseismic modelling of star on the upper main sequence. We also note that for denser stars with masses comparable to the Sun the contribution is negligible which confirms the findings of \cite{2009guzik}.

%sun theoretical paper: villante 2010

%other uncertainties due to opacity smoothing: AAP paper
%opacity interpolation: Yang \& Li 2008: 'Asteroseismic constraints on the OPAL opacity interpolation'

%\subsubsection{New opacity calculations and experiments}

The differences between OP and OPAL concerning the opacity peak temperature of Nickel showed that there are interesting things to be learned from the opacity calculations. The need for a reinvestigation was also realized by atomic physicists. Therefore, new activities in the determination of opacity for astrophysical purposes on both the theoretical and experimental side have started \cite{2011turckchiezeI}.
%,2009whittaker}.
% also
%Pradhan, et al. 2011, 'Re-examination of Stellar Interior Opacities and the Solar Abundances Problem'
%Bailey, J. E.;etal 2009APS..DPPTO6010B	'Laboratory Tests of Stellar Interior Opacity Models'
%2009AIPC.1171...52P, Pradhan, Anil K.; Nahar, Sultana N. 'Accuracy of Stellar Opacities and the Solar Abundance Problem'
%Bailey et al. 2009 Experimental investigation of opacity models for stellar interior, inertial fusion, and high energy density plasmas
The theoretical activities include for example the comparison of spectral opacities for certain elements between different theoretical groups. 
%The OPAS code provides new calculations of opacity including 21 elements and performed by a team at CEA \cite{2011turckchiezeI}
It is important to validate the theoretical results with experimentally determined spectral opacities using modern high-energy laser facilities. These tests are important to check whether the calculations use proper physics.
While the stellar densities are too low to be reproduced in the laboratory, it is possible to draw conclusions by studying equivalent plasma conditions that have similar mean ionization states. New calculations concentrate on the conditions at the base of the solar convection zone and in the driving zone in B stars.
Preliminary results of a comparison of calculated spectral opacities (with 8 codes participating) and experiments done at the LULI 2000 facility in France are given in \cite{2011turckchiezeI,2011turckchiezeII,2011gilles}.
%\cite{2011gilles} compare the theoretical results of Fe and Ni opacity computations with observed spectra obtained at experimental conditions which are relevant to B stars and discuss the differences, e.g. a 'distinct behaviour' of OP. Since this may be due to the computational approach (i.e. physics included) new calculations are currently done to solve these differences.
These experiments/calculations are not only important to determine accurate mean opacities; in fact accurate spectral opacities are very important to determine reliable opacity coefficients for radiative accelerations.

\subsection{Chemical evolution}

Stellar opacities depend on the chemical composition. For example there is observational evidence that the photospheric abundances of certain metal species of B stars may be lower than solar \cite{2009morel} and it was shown \cite{2009montalban} that this chemical composition leads to a higher opacity peak at log~T=5.3 and produces a wider instability strip.
In computations one often assumes the chemical abundances to be homogeneous in the whole envelope. However, atomic diffusion, unless hampered by mixing effects, rearranges elements. 
%mass loss also has effect on driving (bourge comment)

%Consequently, the chemical composition of the stellar envelope changes with evolution. Important processes that govern the chemical stratification are diffusion and mixing.
Atomic diffusion, see e.g. \cite{2009alecian}, is a slow process that modifies the local element abundances due to the counterplay between radiative acceleration vs. gravitional settling which is different for ions of different species. Regrettably, there are currently large uncertainties in the determination of radiative accelerations since they depend on spectral opacities.
In main sequence stars atomic diffusion is responsible for shaping the superficial abundance pattern of Ap and Cp stars. Diffusion may also be partly responsible for the enhancement of opacity around the Z bump in B stars, because elements accumulate due to diffusion where their specific opacity is large.
The effect of diffusion is however swept out if mixing processes are effective.

%chem. evolution of B stars: Bourge, Theado \& Thoul 2007
%lower abundances in B stars: Morel 2006

%jeffery \& Saio 2008: element must accumulate where it's specific opacity is large
%\cite{2007jeffery}: accumulation of nickel also important, and shifts bump to higher Temp. for subluminous B,O stars

%alecian:  diffusion time scales MS: Alecian 2009, in the sun: Turcotte et al. 1998.

%alecian: chemically peculiar stars: sometimes very large overabundances , there are also underabundances
%Am stars (e.g. Sirius A)
%along MS: AmFm, ApBp, HgMn, (non-MS: HB), He

%Delahaye \& Pinsoneault 2006:  comparisons between opac calculations: 50\% differences in radiative acceleration for the iron case in the opacity peak region mentioned in Fig 1 +2 in TC2011a
%TC2011b finds important differences of individual spectra. important for treatment of rad acc.

%sdB P=40-4000 s / 2000 - 9000s (cooler) \\
%kappa mechanism only works if Fe-group elements are levitated and accumulated in layers where kappa mechanism works\\

%GW Vir: kappa mech: L-shell electrons of Fe, CONe-Kshell electrons, red edge: drainage of C/O by gravitational settling.\\

%E.g., diffusion does not seem to have a large influence on asteroseismic modelling of the $\beta$~Cephei star $\theta$~Oph \cite{2007briquet}.

%B star diffusion: 2003 Hempel \& Holweger

Element mixing in main sequence stars occurs due to different processes such as
 convection, convective core overshooting and rotationally induced element mixing like 
 meridional circulation. These processes smooth the stratification of elements
%, angular momentum transport
and our knowledge about their efficiency is still subject to uncertainties.

Among these processes, the extent of overshooting above the convective core is the easiest one to be measured observationally. 
Common values for slowly rotating $\beta$~Cep stars are an overshooting layer with an extent of 0.1--0.4 pressure scale heights \cite{2009thoul}. 
To disentangle the effect of overshooting from rotationally induced mixing, studies of more rapid rotators are needed \cite{2008wadaap}. Asteroseismic analyses of rapid rotators, however, require 2D models which are currently in development \cite{2010reese}.
Nonetheless, the derived overshooting parameter also depends on the chemical stratification and the corresponding opacities. 
\section{Outlook on the near future}

Solving the remaining problems related to mean opacities is important to obtain accurate asteroseismic models. 
%Unsolved problems: \cite{2011balonaIII} find that half of the stars in DSCT strip do not pulsate with amplitudes > 100 ppm.
Precisely determined stellar masses and other fundamental parameters are also important in studies on exoplanets \cite{2011moya}. Satellite missions devoted to the detection of earth-like planets such as {\it Kepler} and CoRoT are currently continuing their observations and provide excellent data for asteroseismic studies. The Canadian mission MOST is also still delivering data despite exceding its projected mission lifetime.

New projects are on the horizon:

BRITE constellation\footnote{http://www.brite-constellation.at} is an Austrian-Polish-Canadian mission consisting of a set of 6 nano-satellites to observe luminosity variations of bright stars. The unique feature of this mission is that the satellites are equipped with a filter in a red or blue passband respectively. The amplitude and phase difference between the two wavelength bands allows for the photometric determination of the surface geometry of pulsation modes at an unprecedent accuracy for bright stars. The launch of the first pair of satellites is scheduled for spring 2012.

The SONG project (short for: Stellar Observations Network Group) consists of a network of 1-meter robotic telescopes devoted to observing bright stars to do asteroseismology and follow-up observations of exoplanet hosts. Each node of this network is equipped with a high-resolution echelle spectrograph.
The prototype node in Tenerife is expected to deliver first light by the end of 2011 \cite{2012song}.

%\subsection{Gaia}
%handler2011: launch foreseen in 2013, provide astrometry, photometry, spectrophotometry and spectroscopy of $\approx 10^9$ objects with 6 $<$ V $<$ 20 . average number of measurements of approx 70 per object

%\subsection{Theory}

Along with additional observations theoretical models and asteroseismic tools are being improved. One example is the open access evolutionary code within the MESA package \cite{2011paxton} which is rapidly developed and currently adapted to asteroseismic use.

Consequently the future of asteroseismology is bright.

% Due to its open access and contribution philosophy may obtain serious improvement.

%\subsection{Convection}

%Muthsam: 
%cepheid Mundpr+Muths2011 even in 2D expensive
%\cite{2011muthsam} ANTARES code: 'imulations in stellar hydrodynamics with rad. transfer and realistic microphysics in 1D, 2D and 3D'
%\cite{2009kupka} Effects of resolution and helium abundance in A star surface convection simulations

%Radek hydrocode

%Houdek

%\section{Appendix}

%M dwarfs: 2011 Rodriguez et al.
%Uytterhoeven et al. (2011): Kepler observations of 750 A-F stars

%high fraction of non-pulsating stars 57.6 < 100 ppm (expected 7-8 due to noise)

%HD 15082, WASP-33 is a star with DSCT pulsations and a massive transiting planet 
%Collier-Cameron et al., 2010): suggest DSCT pulsations, confirmed by \cite{2011herrero},  maybe hybrid star

%Szabo et al 2011

%Balona (2011)
%rotational light variations in Kepler observations of A-type stars

%B star pulsators: also Be stars! fast rotators with emission lines

%\subsubsection{Low frequency modes}

%2011, Lee U.: Amplitudes of low frequency modes in rotating B stars

%\newpage

\end{document}